\documentclass[pre,twocolumn,superscriptaddress]{revtex4}

\usepackage{amsmath}
\usepackage{amssymb}
\usepackage{amsthm}
\usepackage{graphicx}
\usepackage{times}
\usepackage{mathptm}

\theoremstyle{remark}

\theoremstyle{definition}

\begin{document}

\title{Vertex Models and Random Labyrinths: Phase Diagrams for
  Ice-type Vertex Models.}

\author{Kirill Shtengel}
\email[]{shtengel@microsoft.com}
\affiliation{Microsoft Research, One Microsoft Way,
Redmond, WA 98052}
\author{L. Chayes}
\affiliation{Department of Mathematics, University of California,
Los Angeles, CA 90095-1555}

\date{\today }

\begin{abstract}
  We propose a simple geometric recipe for constructing phase diagrams
  for a general class of vertex models obeying the ice rule. The
  disordered phase maps onto the intersecting loop model which is
  interesting in its own right and is related to several other
  statistical mechanical models. This mapping is also useful in
  understanding some ordered phases of these vertex models as they
  correspond to the polymer loop models with cross-links in their vulcanised
  phase.

\end{abstract}


\maketitle
\section{Introduction}
\label{intro} Ice-type models were originally introduced in order
to describe the properties of ice\cite{BF-33:ice,Pauling:ice}, and
have been later generalised to represent other types of
hydrogen-bonded crystals \cite{Slater:KDP}. In ice, the oxygen
atoms form a lattice with the coordination number of four.  Each
bond of the lattice contains a single hydrogen atom that is
shifted from the middle point toward one of the neighbouring
oxygen atoms. It has been conjectured, on the basis of local
electro-neutrality, that precisely two hydrogen atoms are located
near each oxygen atom with the other two being shifted away from
it.  This rule is known as the \emph{ice rule}, and it can be
graphically represented by placing polarisation arrows along the
hydrogen bonds: For each site on the lattice there are exactly two
incoming and two outgoing arrows. There are six such possible
arrow configurations at each site which leads to another common
name for this model (or rather its two-dimensional square lattice
version): the six-vertex model.  Interestingly, the states of
ice-type models can be characterised by topological winding
numbers and the excitations are topological defects carrying
fractional charge \cite{COBN-74:ice}, hence the recent revival of
interest to this type of models in the context of topological
order and fractionalisation in quantum systems \cite{Hermele03a,
Ardonne03,Fradkin04}.  The quantum phase transition in the quantum
square lattice six-vertex model have been argued to mimic the
physics of the HTSC materials with the $d$--density wave order
while their counterparts on the triangular lattice twenty-vertex
model could be relevant to the physics of 2D Wigner
crystals\cite{Chakravarty03}.

The six-vertex model on the square lattice was solved exactly by
Lieb \cite{Lieb-67a,Lieb-67b,Lieb-67c,Lieb-67d} and Sutherland
\cite{Sutherland:6v} using the Bethe Ansatz and also (for some
cases) by mapping it to a soluble free fermion problem
\cite{FW-70}. For a review see \cite{LW:Domb&Green,Baxter}. Much
less is known about the ice-rule models in other contexts.  For
example, in the "next" 2D model, namely the twenty-vertex model on
the triangular lattice, the Bethe Ansatz only works in a few
instances when there are special relations between the parameters
\cite{Baxter-69,Kelland-74a,Kelland-74b}.  More importantly, there
are very few broad-based techniques of general applicability with
which these sorts of models can be studied.

In this paper we develop a new approach to establishing and
understanding the phase diagrams of vertex models with the ice
rule (i.e. equal numbers of incoming and outgoing arrows at each
vertex). We introduce a new class of polymer models which is
closely related to \emph{loop models}, recently a topic of
intensive study (see \cite{VanDerZande} and references therein).
While a lot of progress has been achieved in studying loops models
by the means of exact solutions \cite{BatNW-89:On} and conformal
field theory \cite{Kondev-96:loops}, in the present analysis we
will not recurse to those methods.  The proposed general model is
closely related to so-called \emph{Lorentz lattice gas}
\cite{RC-88:lattice_gas,ZKH-91,BT-92,Cao-Cohen-97} or \emph{random
  labyrinths} \cite{GMV-96} as well as the ice-rule
vertex models as will be discussed below. Such unified graphical
representation on which our approach is based, is not entirely
new, it has been discovered in the context of a loop algorithm
developed to simulate vertex models\cite{ELM-93,Evertz-03}.
Notwithstanding the computational benefits inherent in this
representation, here we will concentrate on the physical insight
into the problem and the theoretical benefits which it provides.
Indeed, our perspective allows for a very intuitive approach to a
general class of vertex models and opens new possibilities for
their rigorous analysis. This approach is ideologically similar to
that of Fortuin and Kasteleyn \cite{FK-72} for the case of the
Potts model.

Since a lot is known about the square lattice six-vertex model,
this is a good reference point from which to start.  We begin by
reviewing a connection between the six-vertex and an intersecting
loop model and show how the phase diagram can be immediately
inferred.  We then apply our method to a particularly interesting
case of the twenty-vertex model identifying its phase boundaries.
A more complete treatment of this problem with necessary proofs
and additional examples shall be presented
elsewhere\cite{CS:unpublished}.

\section{The Switch Model}
\label{sec:switches}

\begin{figure}[htb]
  \begin{center}
    \includegraphics{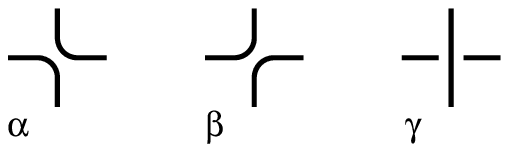}
    \caption{
      The three possible switches for the two-dimensional square
      lattice. We do not distinguish between under- and over-crossings
      for the $gamma$-switches (we are not concerned with topology for
      the purposes of this paper).}
      \label{switches}
  \end{center}
\end{figure}

Let us first introduce a statistical model of objects that we call
\emph{switches}. Consider a finite lattice $\mathbb{L}$ with an even
coordination number. For simplicity let us discuss the case of a
homogeneous lattice and, in particular, postpone the consideration of
boundary sites.  A \emph{switch} is a variable associated with a
lattice site. It is defined as a sorting of all incident edges into
associated pairs.  A more formal definition is as follows:
Consider a complete graph $G(i)$ whose vertices are the bonds $\langle
i,j\rangle$ of $\mathbb{L}$ connected to the lattice site $i$. A
switch $\alpha_k(i)$ is a perfect matching in $G(i)$.
Clearly, if $2m$ is the coordination number, there are $(2m-1)!!$
possible switches for every lattice site. On the square lattice,
the three possible switches, the $\alpha$-switch, the
$\beta$-switch and the $\gamma$-switch are shown in
Fig.~\ref{switches}. In Ref.~\cite{GMV-96}, in the context of
random labyrinths, these have been referred to as NW and NE
mirrors and tunnels respectively..

In general, we shall call the switches the $\alpha_1$-switch, the
$\alpha_2$-switch etc. We will denote by $\mathcal{S}$ a configuration
of such switches on $\mathbb{L}$. The weight of $\mathcal{S}$, which
we denote by $Q\left(\mathcal{S}\right)$ is given as follows: First,
there are \emph{a priori} weights (or activities) associated with
every type of switches. These will be conveniently denoted by
$\alpha_1$, $\alpha_2$ etc.
Next, we observe that $\mathcal{S}$ divides the bonds of the lattice
into loops. Indeed, each bond $\langle i,j\rangle$ is a member of two
switches $\alpha_k(i)$ and $\alpha_{k'}(j)\in \mathcal{S}$ which pair
it with two other bonds etc. We can continue growing this path in both
directions until the loop is closed, i.e. the two "end" bonds are
paired by a switch. A \emph{loop} is thus a cycle of bonds such that
every two adjacent bonds are paired up by a corresponding switch in
$\mathcal{S}$. We are not distinguishing the cycles different only by
a reversal of the overall order. Notice that in general such loop can
visit the same site more than once and in particular, due to tunnels,
can have multiple self-intersections. We let
$\ell\left(\mathcal{S}\right)$ denote the number of such loops and
$\mathrm{A}_1\left(\mathcal{S}\right)$,
$\mathrm{A}_2\left(\mathcal{S}\right)$, \ldots the number of
$\alpha_1$-switches, $\alpha_2$-switches etc. Then $Q$ is given by
\begin{equation}
  Q = \prod_{k=1}^p \alpha_k^{\mathrm{A}_k\left(\mathcal{S}\right)}
  n^{\ell\left(\mathcal{S}\right)}
  \label{Q-weight}
\end{equation}
where $p = (2m -1)!!$ is the total number of distinct switches and $n$
is a positive real number. If $n$ is an integer, it can be thought of
as a number of different colours available for every loop. The
partition function is simply
\begin{equation}
Z \equiv Z_{\alpha_1, \ldots, \alpha_S}=
Z=\sum_{\mathcal{S}} Q\left(\mathcal{S}\right).
  \label{loop_part}
\end{equation}

We also introduce a simple extension of the above switch model in
which we allow two (or more -- depending on the lattice) paths at a
given site to be "fused" together. We shall refer to such objects as
\emph{cross-links}. In general, the cross-links may be \emph{partial}
in the sense that not all pairs of bonds adjacent to a site are fused.
For a four-coordinated lattice such as the 2D square lattice, there
naturally can be only a single type tying up all four adjacent bonds.
Aside from associating the additional \emph{a priori} weight $\phi$
with such cross-links, their introduction results in the following. If
$n$ is an integer, loops that have been fused together are now
required to have the same colour. For an arbitrary $n$ this translates
into the requirement that all such fused loops count as one for the
purposes of evaluating the corresponding weights in
Eq.~(\ref{Q-weight}).

As has already been mentioned, the switch model
and its extension are closely related to several known
statistical-mechanical models.


The $n\to 0$ limit of the switch model describes the Eulerian
walks or cycles (depending on the boundary conditions), i.e. walks
that traverse every bond exactly once.  Self-intersections are
allowed via tunnels. If one were to adopt such a model for
describing a polymer, the weight for the tunnels should translate
into the rigidity of a polymer molecule.


The $n=1$ case corresponds to \emph{Lorentz lattice gas}
\cite{RC-88:lattice_gas,ZKH-91,BT-92,Cao-Cohen-97} or \emph{random
  labyrinths} \cite{GMV-96}. The model in \cite{GMV-96} on a
hypercubic lattice $\mathbb{Z}^d$ is nearly identical to our extended
model, with the following difference: instead of cross-links, there
are \emph{normal sites} at which a path that is otherwise consistent
with the switches can randomly, with equal probability, go in any of
the $2d$ directions. This is equivalent to constructing a random walk
on a graph whose vertices are the cross-links and edges are the
switch-mediated paths between them.  The model corresponds to $n = 1$
since the switches (mirrors and tunnels) as well as cross-links
(normal sites) are distributed with independent probabilities.

For arbitrary $n$, if all switch weights are identical, this is just a
fully-packed limit of the $\text{O}(n)$ loop model \cite{CPS-00}.

A special integrable case of a model (\ref{Q-weight}) with $n<2$
corresponds to a supersymmetric spin chain \cite{MNR-98}

Finally, the $n = 2$ case corresponds to the ice-rule vertex models,
this is the subject of the following section.

\section{Relation to the  ice-type models}
\label{sec:6v}
\begin{figure}[hbt]
  \begin{center}
    { \includegraphics[scale=0.8]{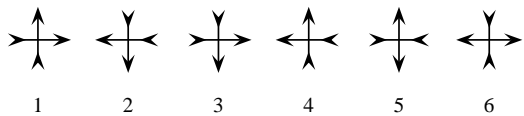} }
    \caption{The six vertices allowed by the ice rule.}
    \label{six-vert}
  \end{center}
\end{figure}

A mapping between $n = 2$ switch models and and ice-type models is
easiest seen in the case of the square lattice six-vertex model whose
possible vertices are shown in Figure~\ref{six-vert}.
Each of those vertices is assigned a corresponding energy $\epsilon_i$
($i=1,\ldots, 6)$.  In the absence of external electric field, the
model should be invariant under the reversal of all arrows thus
requiring $\epsilon_1=\epsilon_2$, $\epsilon_3=\epsilon_4$ and
$\epsilon_5=\epsilon_6$.  For the remainder of this paper we shall
concern ourselves not with the energies as such, but rather with the
(unnormalised) Boltzmann weights $w_i \propto \exp(-\beta
\epsilon_i)$.  Conforming to to the usual notations
(cf.~\cite{LW:Domb&Green,Baxter}), we shall use $a=w_1=w_2$,
$b=w_{3}=w_4$ and $c=w_{5}=w_6$.

\begin{figure}[hbt]
    \includegraphics[scale=0.70]{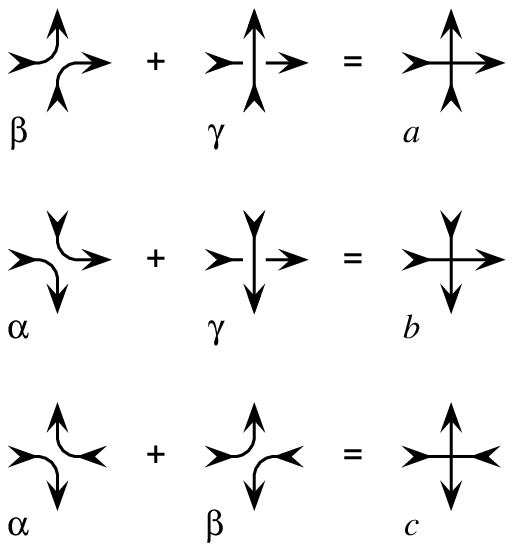}
\hspace{0.8cm}
 \includegraphics[scale=0.75]{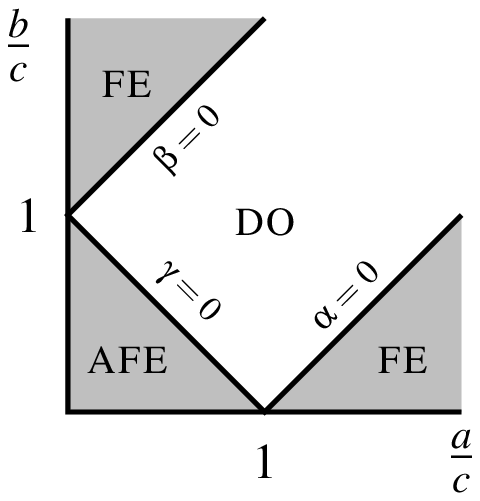}
    \caption{
      (a) Tracing out switches by fusing directed loop segments into
      the ice-type vertices (only types 1, 3 and 5 are shown, types 2,
      4 and 6 are obtained by merely reversing all arrows). (b) The
      phase diagram of the six-vertex model.}
      \label{6-vertex-decomp}
\end{figure}

The mapping of the switch model to the six-vertex model is done by
assigning directions to all loops (since there are two choices for
every loop, the factor of $2^{\ell}$ in Eq.~(\ref{Q-weight}) can be
obtained by summing over all such choices). Tracing out the switch
variables as illustrated in Fig.~\ref{6-vertex-decomp}(a) leaves us
with the six-vertex configurations \footnote{An analogous construction
  for the case of square ice ($a=b=c$) was described in our earlier
  paper \cite{CPS-00}. In fact, similar ideas were previously
  discussed in
  Refs.~\cite{RS:ice,YN:ice,Nienhuis-90:On_square,ELM-93,AN-94,BN-98:6-vertex,Evertz-03},
  but with the exception of \cite{Nienhuis-90:On_square,AN-94}), they
  have not been used as a basis for \emph{analytical} approach.}.
The weights are given by
\begin{eqnarray}
  a = \beta + \gamma ,
\quad
  b = \alpha + \gamma , \quad
  c = \alpha + \beta
  \label{6v-switches}
\end{eqnarray}

In general, we claim that the partition function for the switch model
with weights (\ref{Q-weight}) where $n=2$ is exactly the partition
function for an ice-rule vertex model with vertex weights determined
by
\begin{equation}
  w_i=\sum_{k} \alpha_k \Delta_i^k
  \label{eq:vertex_weight}
\end{equation}
where $\Delta_i^k=1$ if the vertex ${V}_i$ and the switch $S_k$ are
\emph{consistent} and is 0 otherwise.  ${V}_i$ and ${S}_k$ are
\emph{consistent} if there is one incoming and one outgoing arrow for
each pairing of the switch.  (Notice that this automatically preserves
the ice rule.)  The most direct way to see this is to consider a joint
measure on both a switch configuration $\mathcal{S}$ and an arrow
configuration $\mathcal{A}$ with the weight
\begin{equation}
  \tilde Q = \prod_{k=1}^p
  \alpha_k^{\mathrm{A}_k\left(\mathcal{S}\right)}
  \Delta{\left(\mathcal{S},\mathcal{A}\right)}
  \label{eq:joint}
\end{equation}
where $\Delta{\left(\mathcal{S},\mathcal{A}\right)}=1$ if
$\mathcal{S}$ and $\mathcal{A}$ are consistent in the above sense and
0 otherwise.  Now, summing this weight over all possible arrow
configurations $\mathcal{A}$ leads to the the weight of $\mathcal{S}$
given by Eq.~(\ref{Q-weight}) while tracing out switches leaves us
with ice-type arrow configurations where the weight for a single
vertex is given by Eq.~(\ref{eq:vertex_weight}).

We now take a closer look at Eqs.~(\ref{6v-switches}) and notice that
if $\alpha,\beta,\gamma > 0$ (and the switch model only makes sense
for non-negative weights), the allowed values of $a$, $b$ and $c$ lie
precisely within the entire disordered (DO) region of the six-vertex
model (Fig.~\ref{6-vertex-decomp}(b)). In fact, we argue, this is not
a coincidence but a general case: a switch model with \emph{all}
switch weights positive always maps onto a DO phase of a vertex model
\footnote{This in never a truly disordered phase though; due to
  conserved ``arrow currents'', the correlations are always expected
  to be critical (even in higher dimensions) as is the case for the
  six-vertex model.}.  But before we present our argument, we should
discuss the boundary conditions which are defined as follows: let us
regard our lattice $\mathbb{L}$ as actually sitting inside the larger
(possibly infinite) $\Lambda$. Let $\partial\mathbb{L}$ denote those
bonds that connect a site in $\mathbb{L}$ with a site outside. A
boundary condition is defined as a constraint on these bonds' colours
(for $n$ integer) or their external connectedness (general $n$).  From
the perspective of the partition function, we note that the paths in
$\mathbb{L}$ either form closed loops or end at the boundary; each
boundary bond now serves as a ``source/sink'' for a path.

There are several rules that can be now invoked: The simplest two --
which are the extreme cases -- are (i) all such paths count as
additional loops in Eq.~(\ref{Q-weight}) for the weight of a given
switch configuration, or (ii) the only loops counted are the internal
loops.  There are also many intermediate sets of rules
as well as superpositions of various boundary conditions, each with
their own set of rules. Both free and periodic boundary conditions are
examples of such superpositions.  It is noted that none of the above
amounts to any internal constraints -- they are all ``external
constraints'', i.e. all internal switch configurations $\mathcal{S}$
are legitimate.  This is in sharp contrast to boundary conditions that
are imposed by placement of arrows. Indeed, consider the six-vertex
model on a cylinder of girth $L$.  If $m$ upward arrows and $L-m$
downward arrows are placed on the bottom boundary, then the only
allowed switch configurations are those in which there are
$\left|{L-2m}\right|$ paths that start at the bottom and end at the
top.  Under such circumstances, it is easy to see that boundary
conditions are capable of affecting the bulk free energy.

What about the order parameter for the switch model? For an integral
$n$, the natural choice is the expectation value for a bond at the
origin to be of a particular colour. Similarly to the
Fortuin--Kasteleyn (FK) random cluster model, this translates into the
probability that a given bond at the origin
is \emph{connected} to the boundary; with this definition $n$ needs
not remain an integer. Two bonds are considered connected if and only
if there exist a switch-mediated path of bonds between them (using
cross-links is also allowed in constructing such path).  In the
absence of cross-links, this means that the two bonds belong to the
same loop. If cross-links are allowed, the two bonds must belong to
the same cluster of fused together loops.

However, there is an important difference between the order parameter
in the FK model and that in our model. While in the integer $q$ FK
model, the probability that the origin is connected to the boundary is
(at least for the wired boundary conditions) equal to the spontaneous
magnetisation of the corresponding Potts model \cite{ACCN-88}, in our
model such relation between the graphical and the physical order
parameters becomes more complicated. To illustrate this point,
consider again the six-vertex model. Unlike the Ising magnet, it
actually has two ordered phases, ferroelectric (FE) and
antiferroelectric (AFE) that are \emph{not} related to each other by a
simple gauge transformation. While for each given instance of the
six-vertex model only one of these ordered phases can be accessed by
the means of lowering temperature (which we have absorbed in the
vertex weights), there are two possible order parameters with no
obvious relation between them. One of them, namely the uniform
polarisation is discussed in Appendix~\ref{sec:appendix1}. For now it
suffices to note that connectedness to the boundary is a necessary, if
not sufficient condition for any type of physical order; the internal
loops can be reversed at will and thus do not contribute to a
long-range order. Along the same lines, we note that any two locally
defined quantities at two locations can be correlated only if the
corresponding bonds belong to the same loop.  The nature of such
correlation is trivial in terms of colours but may be much more
complicated in terms of arrows. In this sense, the probability of two
bonds to belong to the same loop is an upper bound for any (truncated)
correlation function.

Let us now present a physical (albeit non-rigorous) argument
supporting our earlier statement that a $n=2$ switch model with all
switch weights positive is always in a DO phase. All we have to show
is that less than a finite fraction of bonds in the interior are
connected to the boundary. We start by considering a $n=1$ model in an
upper half-space. Because $n=1$, we do not need to count loops and may
construct the partition function by ``growing'' our system upward from
the boundary layer-by-layer,
the choice for each switch being completely independent of others.
In such a setting it is easy to see that the assumption of a
finite limit of density of (vertical) bonds connected to the
bottom is inconsistent with the finite probability of terminating
any two such strands with finite separation between them (by
turning paths toward each other with the analogs of $\alpha$- and
$\beta$-switches and then connecting them via a finite chain of
tunnels) at every given level. This argument does not preclude a
power-law or even logarithmic decay of such density away from the
boundary, and in fact this is precisely what has been observed
numerically \cite{ZKH-91}. What about $n=2$, the case of our
primary interest? Lacking a proof (the previous part of the
argument can be made rigorous \cite{CS:unpublished}), one can
nevertheless argue that the $n>1$ case always favours
\emph{shorter} loops thus strengthening the above conclusion
\footnote{We should mention here that at least in 2D, such loop
model might still order if $n$ becomes sufficiently large: in our
earlier paper \cite{CPS-00} we actually proved this for a wide
class of models that include the switch model on the square
lattice with $\alpha=\beta$. This ordering, however, happens not
because the loops become long but rather because they become
extremely short. When the entropy associated with large $n$ forces
most of the loops to become of length 4, they break the
translational symmetry of the lattice by circling plaquettes in a
checkerboard pattern. Naturally, such transition is not associated
with any colour symmetry breaking.}. Notice that there is nothing
particularly 2D about this argument, we expect our criterion to
work in any dimensionality.

Let us now turn to the ordered phases. As follows from the above
discussion, a long-range order can not appear unless one of the switch
weights vanishes. To see that this condition indeed correctly
describes the boundary, it is instructive to look at the six-vertex
model again. Let us start with the FE phase boundary, e.g. the one
described by $b=a+c$ or, in our language, $\beta=0$. Notice, that a
line going up can now either continue going up passing through
$\gamma$-switches or turn to the left via an $\alpha$-switch, upon
which it will continue to the left until the next $\alpha$-switch will
send it up again. It can never turn back on itself! Thus, the total
number of loops $\ell$ in Eq.~(\ref{Q-weight}) in an $L \times L$
sample is at most $2L$ and scales sublinearly with the size of the
system $N=L^2$.  As a result, in the thermodynamic limit, the factor
of $2^\ell$ in Eq.~(\ref{Q-weight}) is inconsequential and can be
dropped. Then the partition function is trivially calculated as
\begin{equation}
  Z=\sum_{\mathcal{S}\in\{\alpha,\gamma\}^N}\alpha^{A}\gamma^{N-A}
  =(\alpha+\gamma)^N = b^N
  \label{eq:frozen}
\end{equation}
meaning that the system becomes fully frozen: ether all vertices
are of type 3 or of type 4. Notice that once again, this argument
is not specifically 2D and could be easily used in higher
dimensions, e.g. it generalises the proof by Nagle
\cite{Nagle-69:KDP} of a first order transition in a KDP model on
d-dimensional tetrahedral lattices. Inside this phase, i.e. when
$b>a+c$, we need to extend the model by allowing cross-links so
that $b=\alpha + \gamma + \phi$, $a=\gamma$, $c=\alpha$ (this is
in fact very similar to ``freezing'' a vertex in the loop
algorithm of \cite{ELM-93}). Notice that in general a cross-linked
model is not automatically equivalent to an ice-type vertex model.
One has to verify that a consistent arrow assignment can be done
for all cross-linked clusters with the cross-links corresponding
to the excessive vertices. For the FE phase, cross-linking the
switch model is automatic, but not particularly enlightening since
the model fully orders already at the transition point.  The AFE
phase presents a more interesting case. At the boundary of the AFE
phase $\gamma=0$, i.e.  loops turn at every step and cannot
self-intersect. Assigning directions renders each such loop
consistent with a perfect staggered AFE order, but since each loop
is free to choose one of the two possible directions, no overall
order results. Moving into the phase, we once again resort to
cross-links to compensate for excessive $c$:
$c=\alpha+\beta+\phi$, $a=\beta$, $b=\alpha$. Cross-links do not
change the geometry of loops but rather ``vulcanise'' them into
clusters. It is easy to verify that such clusters are always
consistent with one of the two possible staggered polarisations.
What is essential here is the fact that in the DO phase the loops
are critical, as expected in this type of models on general
grounds.  Therefore such vulcanisation leads to percolation of a
particular colour or, alternatively, staggered polarisation
throughout the sample. What is a bit puzzling, is that the AFE
transition is actually of the infinite order, not of the second
order as our naive argument would generally suggest; this must be
due to the criticality of the phase on the DO side but we do not
have a clearer picture for that in terms of loops.

\section{Twenty--vertex model on the triangular lattice}

So far we dealt with the lattice of coordination number four. This
is somewhat special since the number of possible vertices (up to
the reversal of all arrows) allowed by the ice rule is the same as
the number of possible switches -- three. This is not generally
the case. Let us turn our attention to the triangular lattice. Its
coordination number is six which means there are ${6\choose {3}} =
20$ distinct ways of arranging tree inward and tree outward
pointing arrows. By requiring that the weights for the vertices
are arrow--reversal invariant (zero electric field condition), the
number of parameters is reduced to 10. On the other hand, there
are $5!!=15$ distinct switches. Therefore
Eqs.~(\ref{eq:vertex_weight}) form an underdetermined system. So,
at a first glance the situation appears hopeless: we cannot even
uniquely solve this system, so how could we expect to deduce the
phase boundaries? Nevertheless, the proposed criterion stands: as
long as it is \emph{possible} to satisfy
Eqs.~(\ref{eq:vertex_weight}) with positive switch weights, the
system is in its (critical) DO phase.
\begin{figure}[hbt]
  \begin{center}
    \includegraphics[scale=0.42]{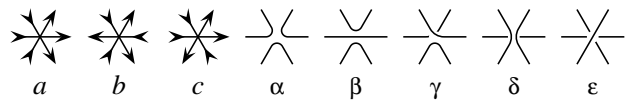}
    \caption{The distinct (up to rotations and reflections) vertices
      and switches on the triangular lattice with their weights. The
      third vertex is chiral, with only one chirality shown; both have
      the same weight.}
    \label{20-vertex}
  \end{center}
\end{figure}
To illustrate this, consider the symmetric version of the
twenty-vertex model (dubbed the F-model in \cite{Baxter-69}) by
requiring that all vertex weights are invariant under the symmetry
group $\text{D}_6$. This leaves us with only three distinct types
of vertices and five types of switches depicted in
Fig.~\ref{20-vertex}. Eqs.~(\ref{eq:vertex_weight}) read:
  \begin{eqnarray}
    \label{eqs:twenty_v_a}
    a=\beta + 2 \gamma + 2\delta + \epsilon,
    \nonumber
    \\
    b = 2 \alpha + 3 \beta + \epsilon,
    \\
    c = \alpha + \beta + 3 \gamma + \delta,
    \nonumber
  \end{eqnarray}
 or, equivalently
  \begin{eqnarray}
    \label{eqs:twenty_v_b}
    a+2c = b + 8 \gamma + 4 \delta,
    \nonumber
    \\
    b+2c = a + 4(\alpha + \beta + \gamma),
    \\
    b+3a = 2c + 4(\beta + \delta + \epsilon).
    \nonumber
  \end{eqnarray}
From the latter form, it is clear that there are no
solutions with all switch weights positive unless $a+2c>b$, $b+2c
> a$, $b+3a > 2c$; according to our criterion this is the DO
region of the model. The suspected phase boundaries correspond to
$\alpha=\beta=\gamma=0$, $\gamma=\delta$=0 and
$\beta=\delta=\epsilon=0$.  These conditions have a simple
geometric interpretation in terms of loops: in the first case,
connections of bonds at $60^0$ are excluded, in the second, the
are no connections at $120^0$ and in the third, there are no
direct tunnels.
\begin{figure}[hbt]
    \includegraphics[scale=0.67]{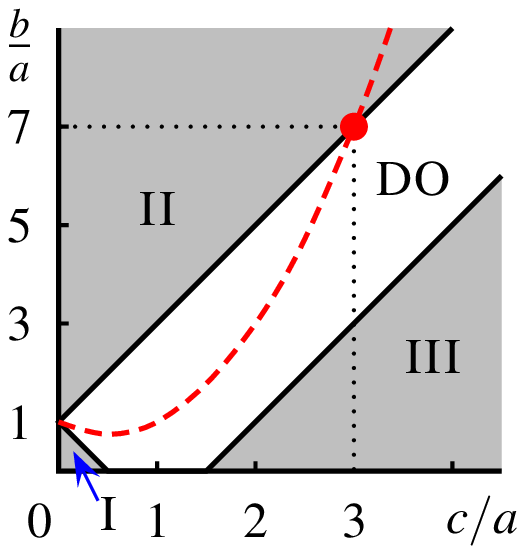}
\hspace{0.2cm}
 \includegraphics[scale=0.53]{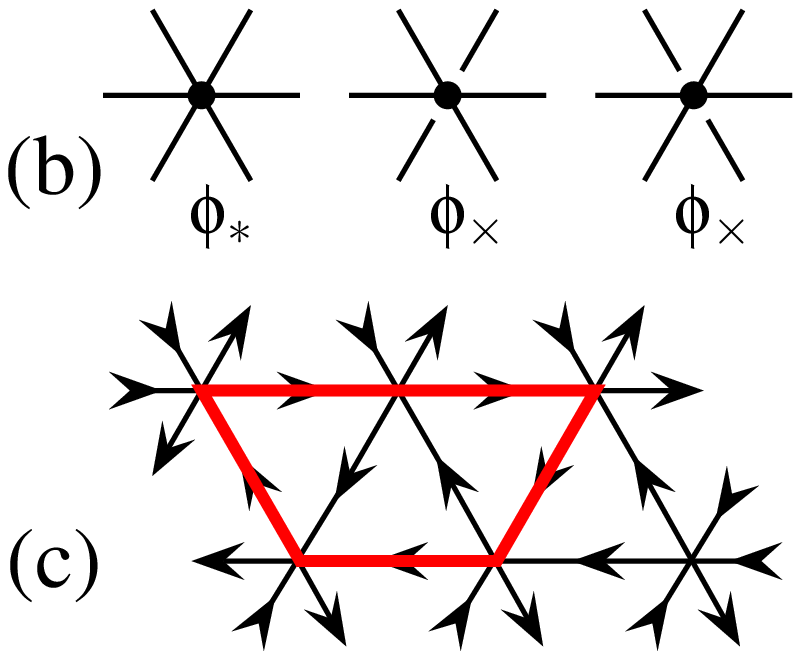}
    \caption{
      (a) The proposed phase diagram for the F-model on the triangular
      lattice shown. Along the dashed line the Bethe Ansatz solution
      is available with the known transition point circled. (b)
      Complete and partial cross-links important for understanding the
      ordered phases II and I (see text for details). (c) A zero mode
      in a state composed entirely of $c$-vertices; the arrows along
      the selected pentagon can be freely reversed.}
      \label{F-phase}
\end{figure}
We now turn to the suspected ordered phases. The simplest one is
the phase II ($b>a+2c$); it is easy to verify that the
introduction of complete cross-links ($\phi_\ast$ in
Fig.~\ref{F-phase}b) corresponding to excessive $b$-vertices is
consistent with the onset of the spontaneous sextipole electric
moment -- a triangular flux phase. This is the only ordered phase
of the model accessed by the means of exact
solutions\cite{Baxter-69,LW:Domb&Green} available along the dashed
line in Fig.~\ref{F-phase}b, a transition to the ordered phase was
found to be of the infinite order.

In the phase I ($a>b+2c$), on the other hand, complete cross-links
are \emph{inconsistent} with the dipolar order described by the
proliferation of $a$-vertices; such cross-link corresponding to
any particular $a$-vertex cannot be broken into all three
different $\delta$-switches.  Physically, this inconsistency
arises from the fact that the symmetry of the ordered phase (mod
complete arrow reversal) is \emph{lower} than the symmetry of the
cross-link and switch weights. One possible remedy is to replace
complete cross-links with the partial ones ($\phi_\times$ in
Fig.~\ref{F-phase}b) breaking such symmetry by hand. The
particular choice of two out of three possible partial switches
prefers a horizontal dipolar order: $a_{-}=2\delta+ \epsilon +
2\phi_\times$, $a_{\diagdown}=a_{\diagup} =2\delta+ \epsilon +
\phi_\times$ where the subscripts for $a$-vertices indicate
directions of their middle out-arrows.  Now once again,
vulcanisation of loops leads to the onset of the dipolar order in
the vertex model, but this is not the original F-model, we have
reduced the symmetry of the $a$-vertex weights. We nevertheless
expect a similar transition to occur in the F-model since the two
models are identical at the transition line of the modified model
and their ordered phases have the same broken symmetry \footnote{
The fully ordered phase of the F-model can have at most six
non-collinear domain walls, meaning there will be macroscopic
regions with dipolar order. Alternatively, it can have an
unlimited number of collinear degrees of freedom: just think of a
state in which all vertices correspond to the same choice of one
of the two partial cross-links in Fig.~\ref{F-phase}b. After  the
cross-linked cluster is assigned directions (so that outgoing
arrows are at $60^0$ to each other), we can chose directions for
the unlinked lines along the third axis independently at random.
Notice, however, that there are at most $L$ such lines (for free
boundary conditions) and therefore the ground state has zero
entropy per site in the thermodynamic limit. It also has a
long-range order with broken dipolar symmetry due to the
cross-linked cluster.}. A plausible counter-argument would point
at the criticality of the DO phase making the above reasoning
suspect: moving in the ``wrong'' direction on the phase diagram
can easily order a critical phase while the system might remain
disordered if we carefully tread in the ``right'' direction. We
argue, this is not the case: the disordered phase here actually
expected to have \emph{finite} susceptibility to the uniform
electric field as is the case for the six vertex model
\cite{LW:Domb&Green,Baxter-70} \footnote{ This is due to the fact
that closed loops, albeit critically long, do \emph{not} couple to
  the uniform field -- see Appendix \ref{sec:appendix1}. The only
  contribution to the susceptibility must come from finitely many
  paths connecting the opposite boundaries of the sample. Notice that
  the existence of a finite number of such paths does not contradict
  our earlier argument for the DO phase which precluded
  \emph{a finite density} of such paths.}.
    Therefore the non-analytic onset of the dipolar order in
any direction of the phase diagram signifies a true phase transition.

Finally, there is Region III ($2c > b+3a$) described by the abundance
of $c$-vertices.  Cross-linking the loops does not appear to work here
since unlike in the previous two cases, some ground states of system
have \emph{local} zero modes (Fig.~\ref{F-phase}c).  What is even more
unsettling, is that in terms of loops a particular zero mode depicted
in Fig.~\ref{F-phase}c contains a direct tunnel, something that
disappears on the boundary of Region III but seemingly reappears well
inside. We therefore can only speculate on the nature (and the
existence) of the transition into this phase, although the fact that a
certain class of loops that could freely fluctuate in the DO phase
disappears at the boundary remains very suggestive.  One possibility
is the onset of an entropically stabilised phase, i.e. an
order-by-disorder transition; further investigation is needed to
clarify this issue.

\section{Twenty--vertex model on the cubic lattice}
Interestingly, there is a one-to-one correspondence between the
vertices (and corresponding switches) in twenty-vertex models on the
2D triangular and 3D cubic lattice (it becomes obvious if one
identifies $\pm 60^0$ directions in Fig.~\ref{20-vertex} with $y$- and
$z$-axes). Reducing the cubic lattice model to its F-model version by
requiring that all vertex weights are invariant under $\text{O}_h$
(the point symmetry group of the cubic lattice), we are left with only
two vertex weights $a$ and $c$ (in notations of Fig.~\ref{20-vertex},
$b\equiv a$) and three switch weights $\alpha$, $\beta$ and $\epsilon$
($\gamma\equiv \alpha$ and $\delta \equiv \beta$). It then follows
from Eqs.~(\ref{eq:vertex_weight}) that $2a = c + 4\beta + 2\epsilon$.
If $2a>c$, the model is in the DO phase. At $2a=c$ we have
$\beta=\epsilon=0$ meaning that loops must turn at every step, this is
a 3D analog of non-self-intersecting loops. Every loop is perfectly
AFE-ordered, and as we have seen, on a 2D square lattice this
signifies a transition to the AFE phase. There is also a general
statement about self-intersections being a relevant perturbation in a
similar type of 2D loop models \cite{JRS-03} suggesting that the
non-intersecting fully-packed loops characterise an unstable fixed
point. So once again, it is suggestive that $2a=c$ is a transition
point.  In 3D, however, the region with $2a<c$ is analogous to Region
III of a triangular lattice model. In particular, there are zero modes
in some of its ground states similar to that shown in in
Fig.~\ref{F-phase}c (naturally, these modes corresponds to the loops
of even length now). Thus, sorting out the properties of Region III of
the F-model on the triangular lattice might shed some light on the
properties of the cubic F-model.

\section{Conclusion}
To conclude, we introduced a class of intersecting polymer loop models
with the loop fugacity $n$. For $n=2$, such a model is equivalent to
an ice-type vertex model in its DO critical phase.  We conjecture that
the boundaries of the DO phase are encountered at the parameter values
which force one or more switch weights to vanish. The onset of a
long-range order in a vertex model is often, but not always equivalent
to the formation of cross-links between polymer loops that results in
a vulcanisation transition.  This correspondence is \emph{not}
strictly 2D, although in this paper we concentrated mostly on 2D
examples to benefit from the additional information available from the
exact solutions. We note that all known transition points in the
ice-rule models satisfy our criterion. As an additional benefit, we
can understand transitions on irregular lattices \footnote{E.g., the
  verticality of the line separating rough from flat phases as a
  function of step repulsion on the phase diagram for the RSOS model
  shown in \cite{RdN-87,dNR-89} becomes clear from our point of view:
  this is just a six-vertex model on a loop-diluted square lattice.
  The transition is determined by a local condition on the vertex
  weights and is independent of the degree of dilution for as long as
  the underlying lattice holds together.}.

However, despite its suggestive nature, it remains a big challenge
to turn our criterion into a rigorous statement.

\begin{acknowledgments}
It is our pleasure to acknowledge helpful discussions with
T.~Prellberg, J.~F.~Nagle, M.~den Nijs, P.~Fendley and C.~Nayak.
\end{acknowledgments}

\appendix
\appendix
\section{Uniform polarisation and susceptibility.}
\label{sec:appendix1}

\begin{figure}[htb]
  \begin{center}
    \includegraphics[scale=0.4]{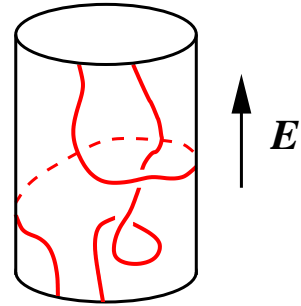}
    \caption{Only paths connecting the top and the bottom of
      the cylinder can be polarised by the electric field parallel to
      its axis. The contribution of each such paths is proportional to
      the cylinder height regardless of the path's geometry.}
    \label{fig:cylider}
  \end{center}
\end{figure}

Consider an $L\times L$ cylinder (or its generalised $d$--dimensional
version -- an $L \times L \times \ldots \times L$ lattice with
periodic boundary conditions imposed in $d-1$ dimensions). The
underlying lattice needs not be square or even regular as long as
appropriate periodic boundary conditions can be imposed. The boundary
condition in the remaining direction, i.e.  along the cylinder axis is
left free.  We now apply a uniform electric field $\mathbf{E}$ along
the axis. The coupling to this field is assumed to be of the standard
form: $\Delta H = - \sum_{\langle ij \rangle}
\mathbf{E}\cdot\mathbf{d_{ij}}$ where $\mathbf{d_{ij}}$ is a vector in
the direction of the arrow along the $\langle ij \rangle$ bond with
the magnitude proportional to the length of the bond (for a regular
lattice it is sufficient to choose all $\left| \mathbf{d_{ij}} \right|
= 1$). Because of the last condition, $\sum_{\langle ij \rangle \in
  \bigcirc} \mathbf{E}\cdot\mathbf{d_{ij}} = 0$ along any closed loop
on the lattice. This allows one to evaluate the partition function of
the vertex model in the form similar to
Eqs.~(\ref{Q-weight}--\ref{loop_part}):
\begin{equation}
  Z =\sum_{\mathcal{S}} \prod_{k=1}^p
  \alpha_k^{\mathrm{A}_k\left(\mathcal{S}\right)}
  2^{\ell\left(\mathcal{S}\right)}
  (2 \cosh \beta E L)^{m\left(\mathcal{S}\right)}
  \label{eq:field_part}
\end{equation}
where $\ell\left(\mathcal{S}\right)$ is the number of closed loops
while ${m\left(\mathcal{S}\right)}$ is the number of paths connecting
the top and the bottom boundaries -- see Fig.~\ref{fig:cylider}.  The
switch weights in Eq.~(\ref{eq:field_part}) correspond to the vertex
weights in the absence of an external field.

The uniform polarisation per site is then
\begin{equation}
  p= \frac{1}{N \beta} \frac{\partial \ln Z}{\partial E} \propto
  L^{1-d}\langle{m}\rangle\tanh \beta E L \xrightarrow[E \to 0]{}
  L^{2-d}  \beta E
  \langle{m}\rangle
  \label{eq:polarisation}
\end{equation}
for small field. Here $\langle{m}\rangle$ is the expected number
of top-to-bottom paths. The uniform susceptibility is then
\begin{equation}
  \chi =  \left. \frac{\partial p}{\partial E} \right |_{E=0}
  \propto L^{2-d} \langle{m}\rangle.
  \label{eq:susceptibility}
\end{equation}

An extensive study of the square lattice six-vertex model in a uniform
electric field presented in Ref.~\cite{LW:Domb&Green} suggests that
this model has a finite non-zero susceptibility to such field
everywhere in the DO region. According to
Eq.~(\ref{eq:susceptibility}) this implies that $\langle{m}\rangle$
remains finite in the thermodynamic limit. The fact that a probability
of a path connecting the opposite sides of an $L\times L$ sample
remains finite as its size grows is consistent with the $r^{-2}$ decay
of the loop correlation function. This is indeed the case for the
points soluble via a free fermion mapping
\cite{Sutherland-68:corr,Baxter-70}. On the other hand, a mapping of
the six-vertex model to an effective Gaussian model (via a restricted
solid-on-solid (RSOS) model) and a Coulomb gas
\cite{Nienhuis:Domb&Green} leads one to think that the exponent
characterising the power law decay of correlations continuously
changes throughout the DO phase. We notice, however, that the
intersecting loops defined by the switch model are not the same as the
non-intersecting loops running along the steps of equal height in the
RSOS model which are described by the "magnetic" exponents in the
Coulomb gas language. It must be the case then that the proper
correlation function for our loops is described by a combination of
magnetic and electric operators "conspiring" to keep its $r^{-2}$
long-distance behavior throughout the DO region.

We finally remark that it has been conjectured in Ref.~\cite{YN:ice}
that an arrow-arrow correlation function for a generic ice-rule model
decays as $r^{-d}$ (although, it seems, the basis for this claim is an
exact result of \cite{Sutherland-68:corr} whose scope is limited to
the free-fermion line of the 2D six-vertex model). This conjecture
appears natural in light of Eq.~(\ref{eq:susceptibility}): such
behaviour is necessary for $\langle{m}\rangle \sim L^{d-2}$ which in
turn is required for a finite non-zero susceptibility to a uniform
electric field.

\bibliographystyle{apsrev}
%

\end{document}